\documentstyle[floats,prb,aps,epsf,twocolumn]{revtex}
\begin{document}
\draft
\title{Andreev reflection in layered structures: implications for \\
high $T_c$ grain boundary Josephson junctions}
\author{Alexander Golubov}
\address{Department of Applied Physics, University of Twente, 7500 AE Enschede, The Netherlands}
\author{Francesco Tafuri}
\address{Dipartimento di Ingegneria dell'Informazione, Seconda Universit\`{a} di Napoli, 81031 Aversa (CE) and ,\\
INFM-Dipartimento Scienze Fisiche dell'Universit\`{a} di Napoli "Federico II", 80125 Napoli (ITALY)}
%\date{\today }
\maketitle

\begin{abstract}
Andreev reflection is investigated in layered anisotropic normal metal /
superconductor (N/S) systems in the case of an energy gap ($\Delta$) in S
not negligible with respect to the Fermi energy (E$_F$), as it probably occurs with
high critical temperature superconductors (HTS). We find that in these
limits retro-reflectivity, which is a fundamental feature of Andreev
reflection, is broken modifying sensitively transport across S/N interfaces.
We discuss the consequences for supercurrents in HTS Josephson junctions 
and for the midgap states in S-N contacts.
\end{abstract}


Andreev reflection (AR) is a scattering process occurring at superconductor/
normal metal (S/N) interfaces that converts an electron incident on a
superconductor into a hole, while a Cooper pair is added to the
superconducting condensate. \cite{andreev} Because of conservation of
momentum, the hole is reflected back in the direction of the incoming
electron and all components of the velocity are substantially inverted if
the exchange momentum in the scattering process is much less than the Fermi
momentum. Retro-reflection occurs whenever the Fermi energy ($E_F$) is much
larger than the gap value ($\Delta $) (Andreev approximation). 
Such an approximation neglects that the retro-reflected hole has in
reality a different momentum $\delta {\bf k}$ in the direction perpendicular
to the S/N interface, which is proportional to the ratio $\Delta /E_F$. \cite
{andreev,been} This means that retro-reflectivity is broken in some
conditions and this will be one of the main issues of this paper.

While predictions based on the Andreev approximation provide accurate
explanations in systems employing low critical temperature superconductors,
in high-critical temperature superconductor (HTS) structures the situation
is more questionable. If we consider that the gap value could be in some
directions of the order of 20 meV (one order of magnitude larger than $%
\Delta $ of traditional superconductors) and that the Fermi energy is
roughly one order of magnitude less than E$_F$ of traditional
superconductors, \cite{kresin} it is interesting to go beyond the Andreev
approximation for systems employing HTS. Concepts based
on AR have been widely used to interpret properties of HTS grain boundary
(GB) Josephson junctions (JJ). \cite{tanaka} Some new interesting arguments
have been developed by taking into account unconventional order parameter
symmetry. Examples are given by the presence of zero bound states in the
density of states of YBa$_2$Cu$_3$O$_7$ in the (110) crystallographic
direction \cite{covin,alff} and more generally by phenomena associated with
broken time reversal symmetry (BTRS). \cite{sigrist}

In this paper we demonstrate that the effects neglected in the Andreev
approximation may determine an extreme depression of Andreev reflection
processes in some directions at S/N interfaces and an enhancement of ordinary 
scattering. The implications of these effects on bound states at interfaces employing
superconductors with d-wave order parameter symmetry are also considered. These 
phenomena can reveal several important features in charge transport in HTS JJ and
enlighten some aspects of the phenomenology of the junctions within the
framework of fundamental issues of HTS, such as anisotropy. We stress that
the effect we consider, being intrinsically related to Andreev reflection,
provides a microscopic explanation of an intrinsic enhanced scattering at
the S/N interface.

Before taking into account an order parameter with a d-wave symmetry
\cite{kirtley}, we consider a layered normal metal
facing an isotropic superconductor with a high value of the order parameter
typical for HTS. This is illustrated in the junction cross section scheme of
Fig.1a and in 3-dimensional view in Fig. 1b. An electron moving along the
planes tilted at an angle $\theta _1$ with respect to the junction interface
is reflected as a hole at an angle $\theta _2$. Locally Andreev reflection
tends to move quasi-particles out of plane and to favor in some way
transport along the c-axis. This counterbalances the fact that quasi-particle
transport in HTS is much favored along the a-b planes. We will present
calculations and phenomenological predictions for the layered structures
reported in Fig.1c and Fig.1d. These can be considered representative of
(100) and non-ideal (001) tilt GB JJs respectively. \cite{mannhart,tafuri}

\begin{figure}
\mbox{\epsfxsize=0.9\hsize \epsffile{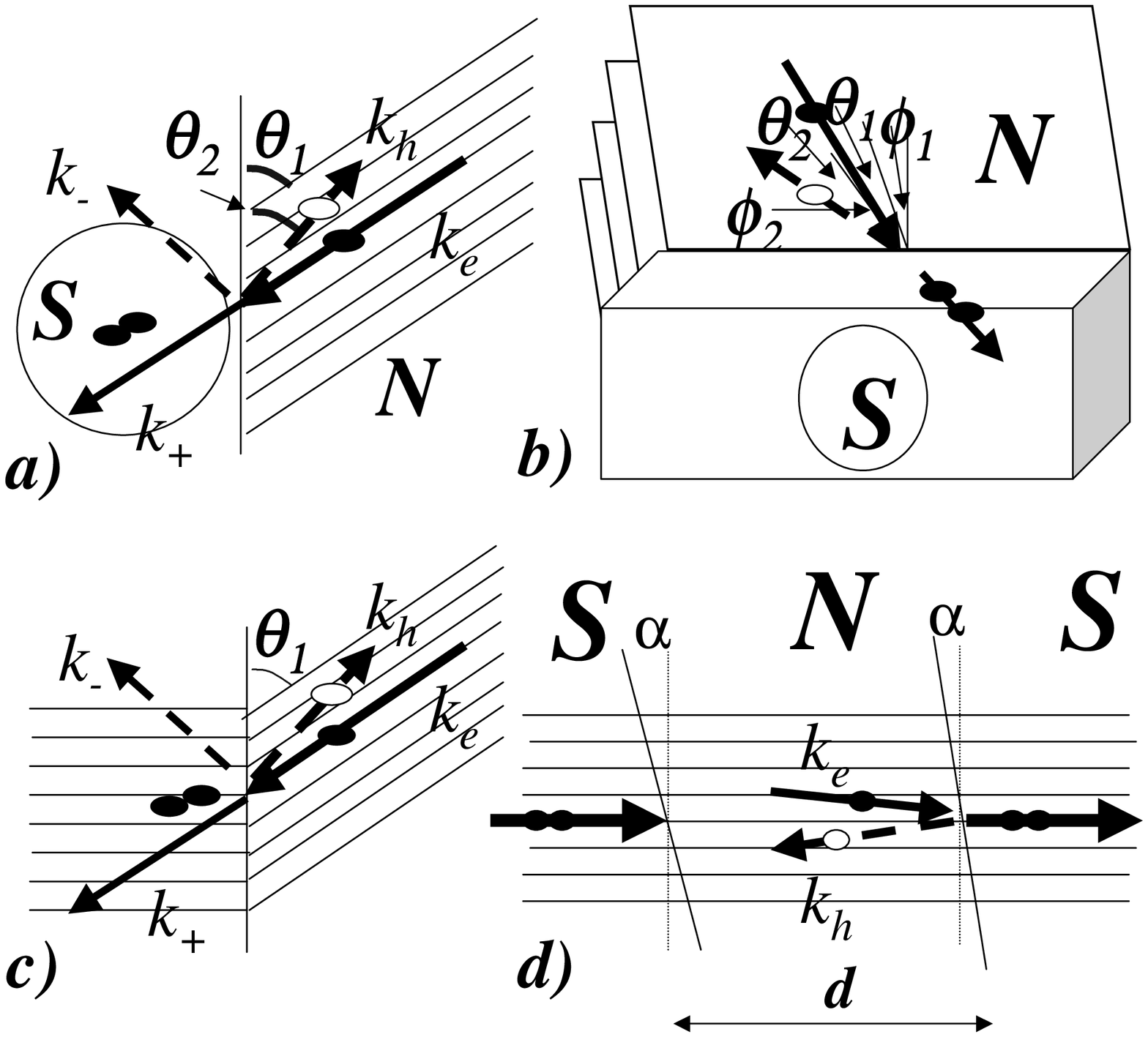}}
\caption
{ a) Cross section of a S/N interface with N layered normal metal and
S isotropic superconductor with a high value of the order parameter. An
electron moving along the planes tilted of an angle $\theta _1$ with respect
to the junction interface is reflected as a hole at an angle $\theta _2$; b)
3-dim. view of (a); c) cross section of a S/N interface like in (a) being
both electrodes layered. d) Scheme of non-ideal (001) GB JJ along with the
supercurrent transport mechanism.}
\end{figure}

{\bf General formalism: Andreev reflection probability for large values of
the }$\Delta {\bf /E}_F${\bf \ ratio.} In order to describe charge transport
through a normal metal - superconductor (N/S) interface, we have used
the Blonder-Tinkham-Klapwijk (BTK) approach \cite{btk} 
introducing some significant modifications. In solving the
Bogoliubov - de Gennes (BdG) equations for the wave functions $\psi _{n,s}$
in the N, S regions, we consider the terms of the order of $\Delta /E_F$
in the expressions of the wave vectors along the incoming (electronic) and
reflected (hole) trajectories in N, $k_{e,h}=k_{Fn}\sqrt{1\pm E/E_F},$ and
along the transmitted trajectories without and with branch crossing in S, $%
k_{\pm }=k_{Fs}\sqrt{1\pm \sqrt{E^2-\left| \Delta _{\pm }\right| ^2}/E_F}$
(see Fig.1a). The expression for $k_{\pm }$ is extended to the general
anisotropic case, when the magnitude $\left| \Delta _{\pm
}^{}\right| $ and the phase $\varphi _{\pm }$ of the gap function in S
are specified on the trajectories $k_{\pm }$. We also generalize the BTK
matching conditions for $\psi _{n,s}$ at the N/S interface: $\psi _n=\psi _s$%
, $\frac{\hbar ^2}{2m}(\psi _s^{^{\prime }}-\psi _n^{^{\prime }})=\left( 
\begin{array}{c}
H_1 \\ 
H_2
\end{array}
\right) \psi (0)$. Here $H_{1,2}=\int (U(x)-E_{e,h})dx$, $U(x)$ is the
interface barrier potential, $E_{e,h}$ are the kinetic energies of the 
electrons and the
holes. In the WKB approximation barrier transmission coefficients $%
D_{e,h}\propto \exp (-2\int \sqrt{U(x)-E_{e,h}}/\hbar dx)$ and $D_h<D_e$
since the hole has less kinetic energy to overcome the barrier. \cite{wolf}
In analogy with the BTK model, the dimensionless barrier strengths for
electrons and holes can be introduced $Z_{1,2}=\sqrt{(1-D_{e,h})/D_{e,h}}$
and generally $Z_1<Z_2.$ This effect may only hold for a realistic extended
barrier \cite{wolf} rather than for a delta-barrier.

With these modifications, we solve the BdG equations and find the
probability of the Andreev and normal reflection process for an electron incoming
from the N side respectively:
\begin{equation}
A(E)=\frac{2(\alpha _1+\alpha _2)^2\beta _1\beta _2\left|
u_{-}^2v_{+}^2\right| }{\left| \gamma _1\gamma _2u_{-}u_{+}e^{i\varphi
_{+}}-\delta _1\delta _2v_{-}v_{+}e^{i\varphi _{-}}\right| ^2},  \label{Andr}
\end{equation}
\begin{equation}
B(E)=\left| \frac{u_{-}u_{+}\gamma _2\eta _1e^{i\varphi
_{+}}-v_{-}v_{+}\delta _1\eta _2e^{i\varphi _{-}}}{\gamma _1\gamma
_2u_{-}u_{+}e^{i\varphi _{+}}-\delta _1\delta _2v_{-}v_{+}e^{i\varphi _{-}}}%
\right| ^2.  \label{Ordin}
\end{equation}
Here $\gamma _{1,2}=\alpha _{1,2}+\beta _{1,2}\mp 2Z_{1,2},\;\delta
_{1,2}=\alpha _{2,1}-\beta _{1,2}\pm 2Z_{1,2},$ $\eta _{1,2}=\alpha
_{1,2}\mp \beta _1\mp 2Z_1,$ $\alpha _1=i\frac{k_{+}}{k_{Fs}},\;\alpha _2=i%
\frac{k_{-}}{k_{Fs}},\;\beta _1=i\frac{k_e}{k_{Fs}},\;\;\beta _2=i\frac{k_h}{%
k_{Fs}},\;$ $u_{\pm }^2=\frac 12(1+i\sqrt{\left| \Delta _{\pm }^2\right| -E^2%
}/E),\quad v_{\pm }^2=\frac 12(1-i\sqrt{\left| \Delta _{\pm }^2\right| -E^2}%
/E)$.

This approach formally presents some analogies
with the formulation of the problem of spin- polarized tunneling in
ferromagnet / superconductor (F/S) junctions. \cite{SF,SF1} A formal analogy
can be established for example between the Fermi momenta in the spin
subbands and the Fermi momenta in the planes $(k_{||}=k_{n+})$ and across
the planes $(k_{\perp }=k_{n-})$ respectively. By increasing the mismatch
between them, in both cases the contribution to the current due to Andreev
reflection is reduced. The effects considered in the present paper
arise from the loss of retroreflectivity due to large gap values, while in
the F/S interfaces they are due to an exchange field in F.

{\bf Josephson current: comparison with HTS JJ}. 
The modification of the probability of the Andreev reflection process
has direct consequences on the calculation of the Josephson current carried by Andreev bound
states (see Fig.1d). Andreev bound states are localized in the barrier region and are
formed by an electron and a hole moving in opposite directions. As a consequence the
momenta mismatch between an electron and a hole leads to a depression of the
Josephson current. As a generic case, we consider tunnel SIS junction, $%
Z_1=Z_2=Z\gg 1$ and both electrodes as s-wave superconductors.

Let an electron have an angle $\alpha _e$ relative to the planes. As
discussed in the introduction, the Andreev reflected hole moves at an angle $%
\alpha _h$ different from $\alpha _e$. The angles $\alpha _{e,h}$ are
related by the conservation of the momenta parallel to the interface $%
sin(\alpha -\alpha _e)k_e=sin(\alpha -\alpha _h)k_h$, where $\alpha =(\pi
/2-\theta _1)$ is the angle between the planes and the interface normal and
the electron (hole) momenta, $k_{e,h}$ are given by $(1\pm
E_B/E_F)k_{e,h}^{-2}=cos^2(\alpha _{e,h}-\alpha )k_{||}^{-2}+sin^2(\alpha
_{e,h}-\alpha )k_{\perp }^{-2}$. Here $E_B$ is the Andreev bound state energy, 
and the anisotropic Fermi surface is approximated by an ellipsoid with axes 
$(k_{\perp }$ , $k_{\parallel })$. The results do not depend qualitatively 
on this choice. 

According to the formalism of Furusaki {\it et al. }\cite{furusaki} the
Josephson current per conductance channel is expressed via the
amplitude of Andreev reflection $a(\varphi ,\omega _n{)}$ in the barrier region (N)
\begin{equation}
I_s=\frac{e\Delta }{2\hbar }\sum_n\left[ \frac{a(\varphi ,\omega _n)}{k_e}-%
\frac{a(-\varphi ,\omega _n)}{k_h}\right] \frac{k_e+k_h}{\sqrt{\omega
_n^2+\Delta ^2}}.  \label{Is}
\end{equation}
Here $\omega _n=\pi T(2n+1)$ and $\varphi $ is the phase difference. The
amplitude $a(\varphi ,\omega _n{)}$ is found from the solution of BdG
equations and describes the multiple scattering in the barrier region. 
In a tunnel junction $E_B=\Delta $. We neglect here a weak energy dependence 
of $\Delta $ due to non-constant density of states in the relevant energy range.

\begin{figure}
\mbox{\epsfxsize=0.8\hsize \epsffile{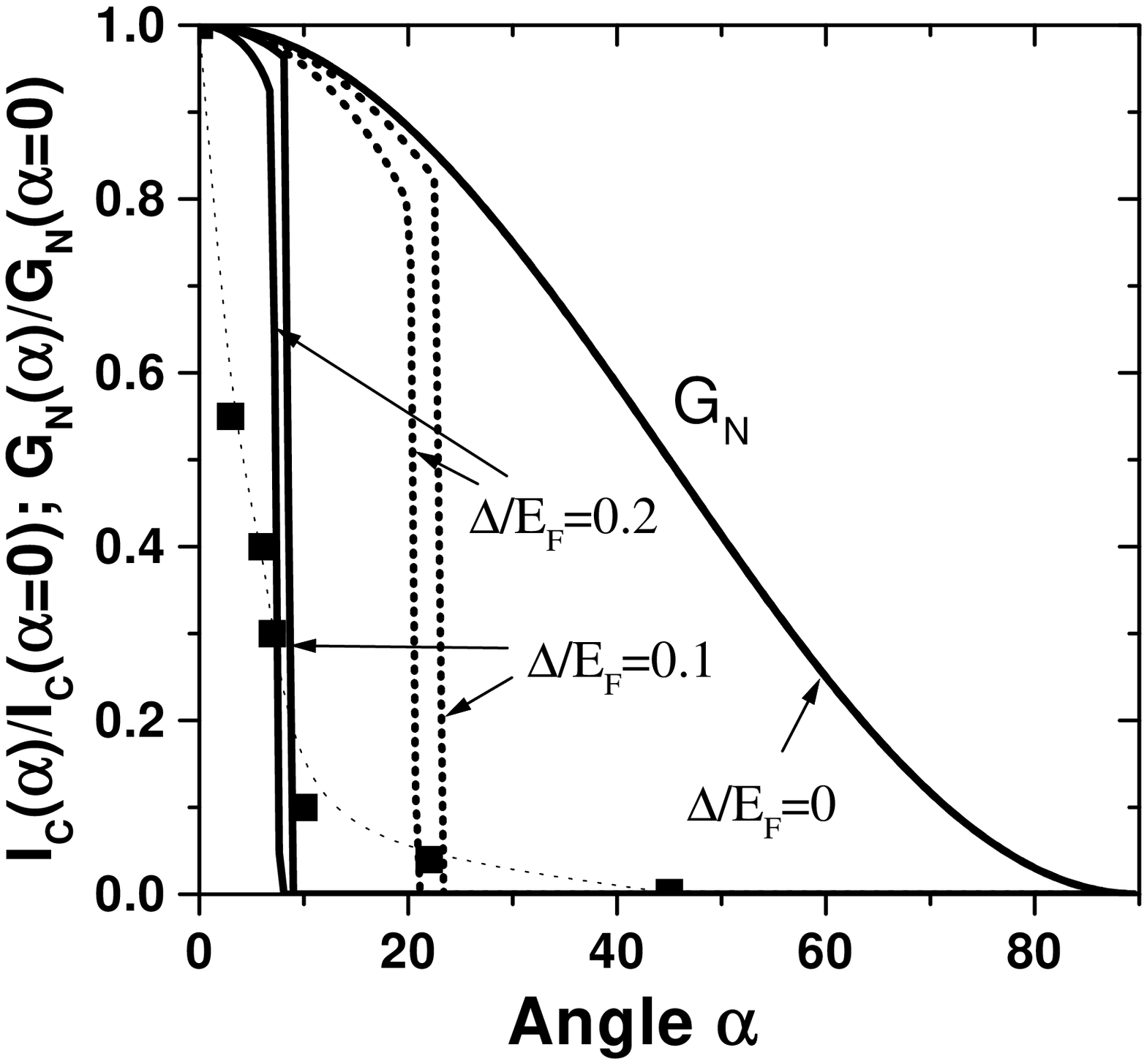}}
\caption
{ Dependence of $I_C$ and $G_N$ on the angle $\alpha $ between the
planes and the interface for different values of the $\Delta /E_F$ ratio and
the Fermi mismatch $(k_{\perp }/k_{\parallel })$=0.1 (solid lines) and 0.3
(dashed lines). Experimental data for (100) tilt grain boundaries are shown
by dots.}
\end{figure}

The angle dependence of $a(\varphi ,\omega _n{)}$ in Eq.(\ref{Is}) is mainly
controlled by the scattering amplitude $t_{eh}$ in the electron-hole
channel, $a\propto t_{eh}(\alpha ,\alpha _e,\alpha _h)$. The normal state
conductance of the junction $G_N$ per channel is determined by the
scattering amplitude $t_{ee}$ in the electron-electron channel $G_N\propto
t_{ee}(\alpha ,\alpha _e)$. The explicit form of $t_{eh},$ $t_{ee}$ depends on
a choice of the shape of a potential barrier. For the $\delta $-barrier
the BdG solution yields $a\propto t_{eh}\propto
k_ek_hk_{Fn}^{-2}(1+Z^2)^{-1}$, $G_N\propto _ek_e^2k_{Fn}^{-2}(1+Z^2)^{-1}$.
As it follows from the above set of equations for $k_{e,h}$, no real
solution exists for $k_h$ for the angles $\alpha _e>\alpha
_e^{th}$, where the Andreev reflection process is prohibited. The threshold angle 
$\alpha _e^{th}$ sensitively depends on $k_{\perp }/k_{\parallel }$ and 
$\Delta /E_F$. 

$I_c$ and $G_N$ are obtained by considering  all 
conductance channels, i.e. integration over the angle $\alpha _e$.
The final result depends on the barrier shape, which controls the relation
between $\alpha _e,\alpha _h$ and trajectories in S regions. We consider
below an extended barrier for strong directional tunneling  
around the normal direction (the tunneling cone effect).\cite{wolf}. 
The results of the numerical calculations of $I_C(\alpha )$ and $%
G_N(\alpha )$ are shown in Fig. 2 for different values of the $\Delta /E_F$
ratio and Fermi mismatch $k_{\perp }/k_{\parallel }$. We give evidence of a
remarkable decrease of $I_C$ with the increase of the angle $\alpha $. The
reason for this drop is the existence of both a threshold angle 
$\alpha _e^{th}$ and a narrow tunneling cone. This result is in qualitative 
agreement with experimental data obtained on GB JJs.\cite{mannhart} An account of the
distribution of tunnel angles in a real interface would broaden the sharp
transitions in Fig. 2.

The general picture considered above can be also applied to the geometry
shown in Fig.1c providing the same qualitative behavior. A rigorous
treatment would require some further hypothesis on Cooper pair
transport in the c-direction that is beyond the scope of this paper.

The interplay between the effect of loss retro-reflectivity considered in this paper
and well established effects such as interface roughness or the co-existence of order parameters 
with different symmetry close to junction interfaces \cite{tanaka,Burk,Barash,GK} 
deserves further investigations. Interface roughness
produces a smooth dependence of $I_C$  on the angle $\alpha$, in contrast to
the sharp dependences shown above.

\begin{figure}
%\begin{center}
\mbox{\epsfxsize=0.8\hsize \epsffile{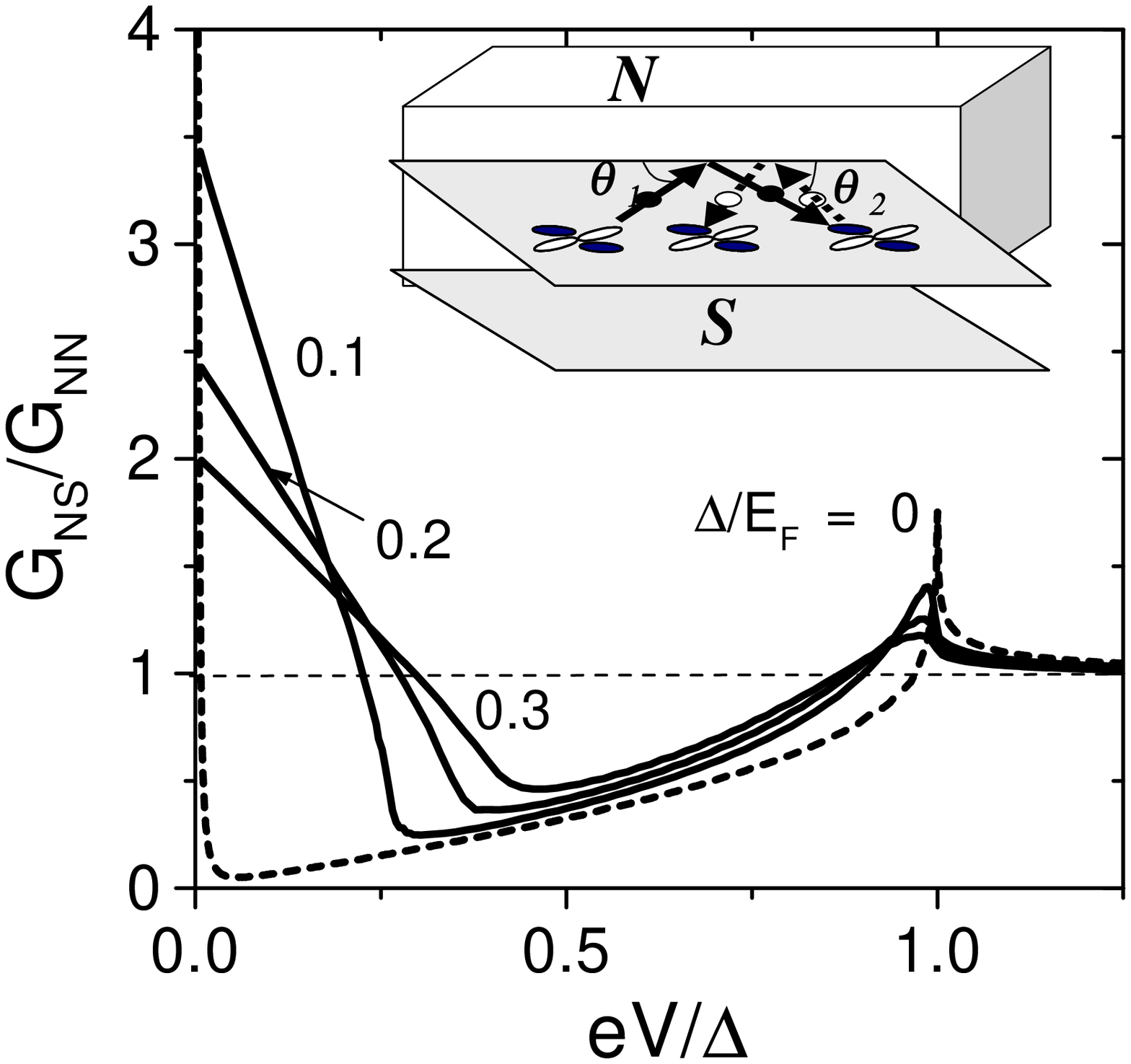}}
%\end{center}
\caption
{ The conductance for different values of the $\Delta /E_F$ ratio in
a tunnel NIS junction. Inset: resonant state originating at the interface of
a d-wave S facing an insulator or a normal metal N' along the (110)
orientation.}
\end{figure}

{\bf S (anisotropic superconductor) / N interface: conductance and the
problem of zero bound state.} We consider another aspect typical of HTS
junctions related to Andreev reflection by introducing a d-wave order
parameter for the S and investigating the origin of zero bound states (ZBS). The
basic process is shown in Fig. 3, where a d-wave S faces an insulator or a
normal metal N' along the (110) orientation (the a-b planes can be in
principle rotated of an angle with respect the S/N interface different from $%
45^{\circ }$). An electron, coming for instance along the direction of the
positive lobe of the order parameter, first suffers an ordinary scattering
at the interface with a normal metal (N') and then is Andreev reflected 
towards the negative lobe of the order parameter. Then the hole
will experience an ordinary scattering at the S/N' interface and will be
reflected towards the positive lobe of the order parameter, where it will be
Andreev reflected again. This process produces a constructive interference
and as a consequence a resonant state, formally described by a pole in the
Andreev amplitude Eq.(\ref{Andr}). This manifests itself as a zero bias peak in
the density of states in S (zero bias anomaly) (ZBA).\cite{covin,Tanaka}

These arguments are essentially based on the retroreflection property of
Andreev reflection. On the other hand the constructive interference breaks down for high
values of $\Delta /E_F$: the electronic state created after two
Andreev and two normal reflections will not propagate along the same
trajectory as the initial one. As a consequence ZBS will be damped and the low voltage
conductance will be decreased. 

\begin{figure}
%\begin{center}
\mbox{\epsfxsize=0.8\hsize \epsffile{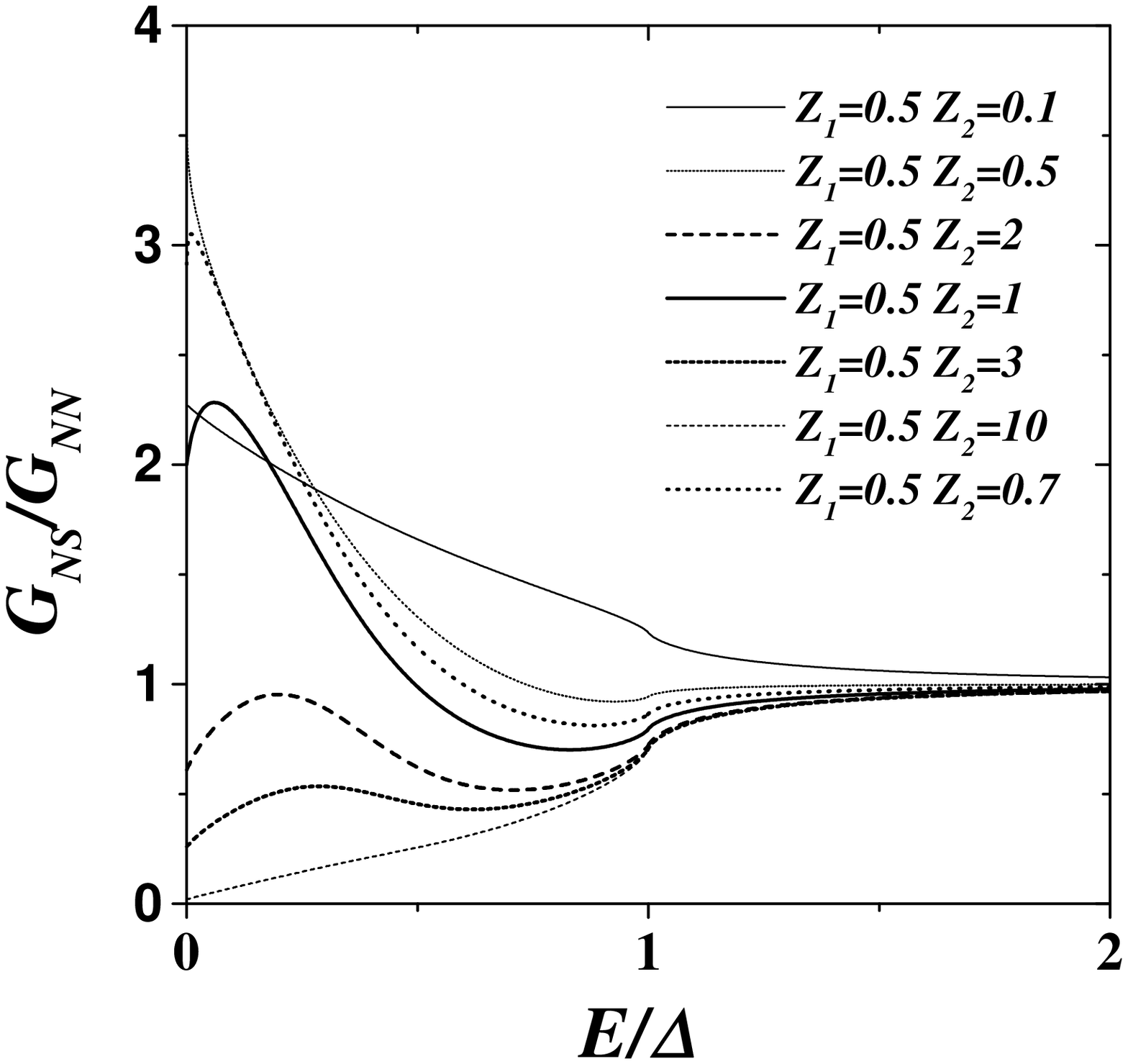}}
%\end{center}
\caption
{ The conductance for fixed value of $Z_1=0.5$ and for different
values of $Z_2$ at temperature T = 0 K. }
\end{figure}

We have calculated the conductance of a tunnel NIS junction ($Z_1=Z_2\gg 1)$
selfconsistently by the method of Ref.\cite{GK}, taking into account the
angle mismatch of electrons and holes and the selfconsistent reduction of the pair
potential at the interface. The angle averaging was performed by choosing a $%
\delta $-function potential barrier with an angular dependence of the
transmission coefficient $D(\theta )=D(0)\cos ^2\theta $. The results of
calculations are presented in Fig.3 for misorientation angle $\alpha =45^0$
and different values of the $\Delta /E_F$ ratio, giving some evidence of broadening
of the ZBA.
We also point out that in contrast with ZBS broadening 
due to surface roughness \cite{Barash,GK} the 
mechanism we propose has an intrinsic nature mainly controlled by the 
$\Delta /E_F$ ratio.

The problem of the ZBA has also been investigated in the case that $Z_1$ is
different from $Z_2$. This difference is particularly meaningful in the
formation of the ZBS, where both electron and hole scattering processes
across the same interface are involved. In Fig. 4 we report the conductance
in the regime $\Delta /E_F=0$ for fixed value of $Z_1=0.5$ and for different
values of $Z_2$ at temperature T = 0 K. We notice the appearance of peaks at
finite voltages and the removal of the ZBA for some values of $Z_2>Z_1$. 
This means
that the crossover from ZBA\ to bound states at finite voltages 
can also take place due to a different
transparency of electrons and holes at the S/N interface. Such a crossover
has been also predicted for ferromagnet -insulator - superconductor
junctions by Kashiwaya {\it et al}. \cite{SF1} Our result is obviously
independent of any ferromagnetic electrode or barrier and only relies on a
possible barrier asymmetry for holes and electrons. Barrier asymmetry acts as a
kind of filter which creates an electron-hole imbalance, thus reducing the
probability of the formation of the ZBA.

In conclusion, Andreev reflection has been theoretically investigated in
layered anisotropic normal metal / superconductor (N/S) systems in the case
of an energy gap ($\Delta $) in S not negligible with respect to the Fermi energy
($E_F$). We have demonstrated that the combination of large gap and strong
anisotropy leads to an intrinsic decrease of the critical current density as
a function of the tilt angle in HTS Josephson junctions, as experimentally
observed. A damping of the resonances originating the zero
bound states has been also found.

{\bf Acknowledgments.} The authors would like to thank A. Barone, Z. Ivanov,
J.Kirtley, M.Yu. Kupriyanov, A. Leggett, J. Mannhart and I.I. Mazin for helpful
discussions. F.T. was supported by the Istituto Nazionale di Fisica della
Materia (INFM) under the project PRA ''High Temperature Superconductor
Devices'' and by a MURST COFIN98 program (Italy).

\end{document}